\documentclass[fleqn,10pt]{wlscirep}
\usepackage{amsmath}
\usepackage{soul}
\usepackage{hyperref} 
\newcommand{\beginsupplement}{%
        \setcounter{table}{0}
        \renewcommand{\thetable}{S\arabic{table}}%
        \setcounter{figure}{0}
        \renewcommand{\thefigure}{S\arabic{figure}}%
     }

\title{Inventor collaboration and its' persistence across European regions}

\author[1,2]{Gergő Tóth}
\author[1,3]{Sándor Juhász}
\author[1,3]{Zoltán Elekes}
\author[1,4,*]{Balázs Lengyel}
\affil[1]{Agglomeration and Social Networks Lendület Research Group, Hungarian Academy of Sciences, 1097 Budapest, Hungary}
\affil[2]{Spatial Dynamics Lab, University College Dublin, Belfield, Dublin 4, Ireland}
\affil[3]{Faculty of Economics and Business Administration, University of Szeged, 6722 Szeged, Hungary}
\affil[4]{International Business School Budapest, 1037 Budapest, Hungary}

\affil[*]{Corresponding author: lengyel.balazs@krtk.mta.hu}


\begin{abstract}
Inventor collaborations that span across regions facilitate the combination and diffusion of innovation-related knowledge. While remote partnerships are gaining importance; most R\&D cooperations remain embedded in local environments and little is known about how spatial patterns of creation and persistence of ties in large-scale co-inventor networks differ. We use the publicly available OECD REGPAT database to construct a co-inventor network of the EU27 and continental EFTA countries from patents filed between 2006 and 2010, and identify those ties that had been persisted from before 2006. A community detection exercise reveals that persisted collaborations cluster at a smaller geographical scale than what is observed in the full network. We find that in general the estimated probability
of persisted collaboration does not differ from the complete collaboration network when geographical distance, technological similarity or the number of common third partners across regions are assumed to influence tie probability separately. However, persistent collaboration gains exceptional likelihood if regions are proximate in all three dimensions. Our results therefore offer evidence that repeated inventor collaboration drives regional innovation systems towards isolation, which is a threat for European innovation policy.
\end{abstract}

\begin{document}

\flushbottom
\maketitle
\thispagestyle{empty}

\section*{Introduction}

\vskip 0.1in
Technological innovation is concentrated in space due to its increasing returns to scale \cite{audretsch1996r, bettencourt2007growth}, and because it is easier to share complex knowledge with partners in geographical proximity and through face-to-face interaction \cite{anselin1997local, balland2017geography, jaffe1993geographic,storper2004buzz}. Social  relations greatly influence these phenomena by increasing externalities \cite{katz1985network}, by facilitating the emergence of novel combinations \cite{acemoglu2016innovation}, and by enabling flows of innovation-related knowledge through connections that can bridge even great distances \cite{breschi2009mobility, sorenson2006complexity}. Due to its importance, the spatial structure and dynamics of innovative and R\&D collaboration networks have been extensively investigated in economic geography and related fields \cite{cantner2017coevolution,cassi2015research, balland2012proximity,brenner2013introduction,fleming2007evolution,fritsch2004innovation,maggioni2007space, ter2013dynamics}.  However, surprisingly little is known about how the spatial patterns of persistent non-local inventor collaborations differ from newly created ones.

\vskip 0.1in
According to a central stylized fact, dense local networks represent most of the interactions related to innovation, while relatively sparse distant connections provide extra opportunities for novel combinations \cite{bathelt2004clusters}. The dynamics of collaboration networks are thought to mirror the evolution of the innovation system itself \cite{gluckler2007economic}, in which a major threat is that network evolution leads to a dead-end in technological development in regions\cite{grabher1993weakness}. Theoretical claims emphasize that such locked-in development is more likely to occur if strong connections, dyadic similarities or triadic network cohesion are too strong drivers of spatial network formation\cite{boschma2010spatial,uzzi1997social}. Empirical findings supports these claims by showing that besides geographical proximity, the overlap of technological portfolios\cite{balland2012proximity,cassi2015research} and triadic closure of partnerships\cite{ter2013dynamics} also increase the probability of cross-regional innovative collaboration.

\vskip 0.1in
To better understand whether spatial network dynamics leads regions to lock-in, Juhász and Lengyel\cite{juhasz2017creation} proposed a simple framework for separating the mechanisms of tie creation and tie persistence. This distinction is important because uncertain new connections offer access to new knowledge, whereas persistent ties represent those strong collaborations that are worth repeating despite their considerable opportunity costs \cite{dahlander2013ties,rivera2010dynamics}. They argue that the sign of the correlation between the probability of tie persistence and the joint effect of dyadic variables (\textit{e.g.} geographical distance, technological similarity, triadic closure) indicates whether the network evolves towards an inert and too cohesive system, or remains open for variety through new connections. A positive correlation then means that network evolution is headed towards lock-in, whereas a negative correlation indicates that some network variation is still present to offset network closure. 

\vskip 0.1in
In this paper, we investigate the spatial patterns of co-inventor collaboration and the persistence of patenting partnership in the regions of the EU27 and continental EFTA countries (Norway and Switzerland). We use the publicly available OECD REGPAT Database that contains the location of inventors at NUTS3 level regions (Nomenclature of Territorial Units for Statistics) defined by the European Statistical Office. The European patenting collaboration has been claimed to remain fragmented\cite{chessa2013europe} by country borders despite the policy efforts to strengthen international collaboration \cite{arrieta2017quantifying} and the fact that international collaboration produces better patents in terms of citations \cite{lengyel2016international}. Our community finding exercise\cite{blondel2008fast} provides further support for this claim, and more importantly, we find that persisted co-inventor ties are concentrated at an even smaller geographical scale. 

\vskip 0.1in
In a further analysis, we investigate how the probability of persistent co-inventor ties, as a function of geographical distance\cite{liben2005geographic,lambiotte2008geographical,lengyel2015geographies}, the overlap between technological portfolios and the number of common third partners and differ from co-patenting in general. Then, we apply a multivariate gravity equation approach\cite{broekel2014modeling, maggioni2007space}, and control for region-level covariates such as population density, gross value added and centrality in the network among others. Because the main explanatory dyadic variables correlate with each other\cite{lambiotte2008geographical}, in the final step, we analyze the joint effect of these variables. Although the patterns of persistent ties do not differ considerably from co-inventorship in general neither in the bivariate nor in the multivariable setting, we find very strong and positive correlation between the threeway interaction term and the probability of persisted ties.  

\vskip 0.1in
Our findings together suggest that persistence of co-inventorship drives European regions toward cohesive and perhaps also locked-in systems of innovation.
\vskip 0.1in

\section*{Results}
To analyze the spatial patterns of inventor collaborations and their persistence across European regions, we use the full set of patents authored by European inventors, and registered by the European Patent Office (EPO). We look at inventor-inventor collaborations on patents that have been registered in the 2006-2010 period, and identify a collaboration as persisted if the inventors had co-authored at least one patent previously, during the 1991-2005 period. The data is from the publicly available OECD REGPAT Database that contains the year of filing and technological classes of patents, and the unique identifier and the region of residence of inventors. Data management is explained in detail in the Methods section. \par 

\vskip 0.1in

\begin{table}[ht]
\centering
\begin{tabular}{lcc}
\hline 
 & \textbf{Collaboration} & \textbf{Persisted Collaboration} \\
\hline
Number of Individual Collaborations & 772,378 & 41,883 \\
Number of Region Ties & 46,857 & 6,200 \\
Density of the Region Network & 0.05 & 0.006 \\
Number of Communities in the Region Network{*} & 7 & 23 \\
Modularity of the Region Network & 0.372 & 0.584 \\
Relative Modularity of the Region Network & 0.329 & 0.379 \\
\hline
\end{tabular}
\caption{\label{tab:Table1}Descriptive characteristics of the co-inventor collaboration network and the persisted collaboration network. {*}Communities of size 1 are excluded. }
\label{table:Table1}
\end{table}

We summarize the major characteristics of cross-regional inventive collaborations in Europe and their persistence in Table\ref{table:Table1}. Out of more than 772K individual cross-regional collaborations of inventors in 2006-2010, only every 18\textsuperscript{th} was repeated from a previous collaboration. We aggregate these individual ties to European NUTS3 regions, creating a weighted network of regions in which the weight of a region-region tie is the number of individual collaborations connecting the two. After the aggregation, the collaboration network of regions consists of nearly 47K ties, while a persisted collaboration exists only between one 7\textsuperscript{th} of these region pairs. Consequently, the density of the persisted network is one magnitude lower than the density of the complete collaboration network.\par

\subsection*{Spatial communities of inventor collaboration}

To characterize the spatial structure of the interregional networks of collaboration and persisted collaboration, we use the Louvain algorithm~\cite{blondel2008fast}, a popular and effective algorithm to partition the network by hierarchical clustering. We then calculate the modularity of the community structure to measure the tendency of connections to be within groups of the partition rather than between them~\cite{newman2006modularity}. The modularity $Q$ of the network's partition can be written as:
\begin{equation}
$$Q = \sum_{k=1}^{K} \Big[ \dfrac{L^{w}_{k}}{L} -\Big(\dfrac{L_{k}}{L}\Big)^{2} \Big]$$,
\end{equation}
where $L$ is the total number of individual co-inventor edges in the network, $L_{k}$ is the total edges of members of group $k$, and $L_{k}^{w}$ is the number of edges within group $k$. Because modularity is highly dependent on the size and density of the network, following Sah et al.~\cite{sah2017unraveling}, we calculate the relative modularity of the two networks by dividing \textit{Q} by the theoretical \textit{Q\textsubscript{max}} that would be achieved if all edges were within the communities. Rand indexes between the presented and four other community structures found on randomly re-shuffled matrices are reported in SI Table 1 and SI Table 2.  

\vskip 0.1in
Our findings reported in Table\ref{tab:Table1} reveal remarkably less communities in the collaboration network, than in the persisted collaboration network. Persisted collaboration tends to be concentrated to a higher extent within these communities than collaboration itself, which is due to the lower level of network density, since relative modularity has similar values in the networks.

\begin{figure}[!hb]
\centering
\includegraphics[width=\linewidth]{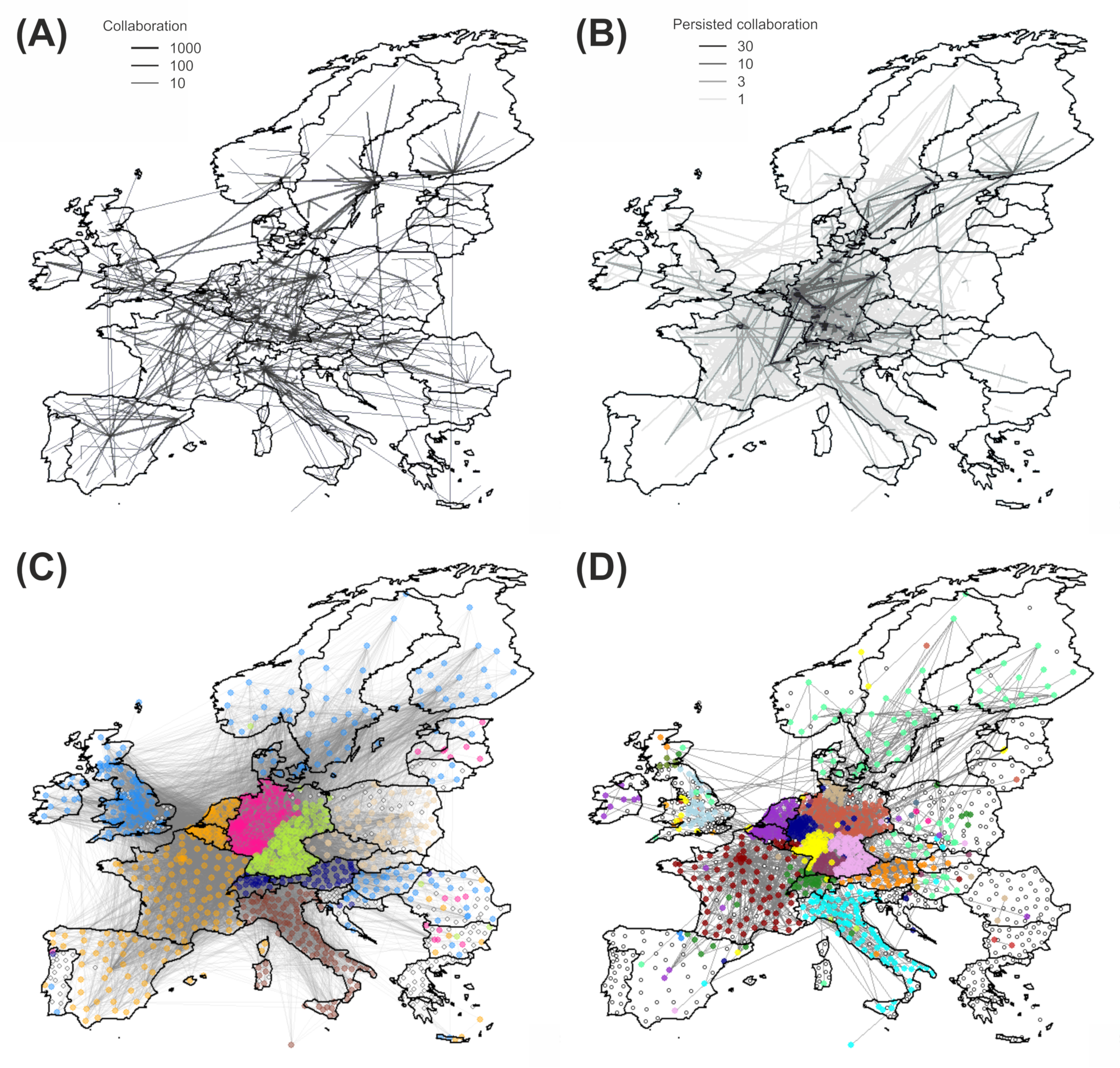}
\caption{Spatial patterns of inventor collaboration and persisted collaboration networks of European regions. \textbf{(A)} The maximum spanning tree of the collaboration network across NUTS3 regions in Europe reveals the importance of national centers. \textbf{(B)} Most of the persisted collaborations remain within country borders, and strongest ties are concentrated within close proximity of innovative hubs. \textbf{(C)} The 7 communities of the collaboration network span across countries, but are mostly concentrated in large regions. \textbf{(D)} Persisted collaboration is organized into 23 smaller-scale clusters.}
\label{fig:Fig1}
\end{figure}

\vskip 0.1in
In Figure\ref{fig:Fig1} we map both networks and their community structures. Because the collaboration network has too many edges to put all on the map, and the edge weights deviate on a large scale, we simplified the network to its' maximum spanning tree for the illustration, in which every region is connected to every other region by only one path such that the sum of the tie weights are maximized. Figure 1A reveals that most European countries have an outstanding innovation center, in which collaboration is concentrated, and these centers bridge the innovation system with other countries (\textit{e.g.} in Spain, Sweden, Finland, Italy, Hungary and Romania). There is more than one center in Poland, France and especially in Germany, where regional centers emerge from the maximum spanning tree. Most of these spatial structures are present in the persisted collaboration network as well, in which all ties are depicted in Figure 1B. Persisted collaboration is concentrated in single innovation hubs in Sweden, Finland, Italy, whereas two centers emerge in France (Paris and Lyon) and multiple centers in Germany, where these centers are strongly connected to each other but also collect even stronger local connections. 

\vskip 0.1in
Both networks are organized into spatially bounded communities (Figure 1C and B), which is not surprising since spatial community structures have been repeatedly found in social and communication networks \cite{lambiotte2008geographical,lengyel2015geographies,sobolevsky2013delineating}. Chessa et al.\cite{chessa2013europe} also report that communities of inventor collaboration are bounded by national borders in Europe. Our findings suggest more integration in the studied period, since some of the communities expand countries: for example the community including the Benelux states, France and Spain. These communities contain neighboring countries such as Switzerland and Austria, or the community of Poland, Czech Republic and Slovakia. However, the United Kingdom, Hungary and Croatia are found to be in one community with the Scandinavian countries. We find that inventor collaboration in Italy is a separate network community. Interestingly, Germany is organized into two large communities and this spatial structure does not follow the traditional East-West divide, which is a sign of reorganization of inventive collaboration after the fall of the Iron Curtain\cite{jun2017meet}.

\vskip 0.1in
Interestingly, the persisted collaboration network is concentrated in smaller spatial communities than the complete collaboration network. The innovation systems of persistent collaboration are not organized by a universal schema of spatial levels. In some cases these systems represent countries (\textit{e.g.} France and Italy), groups of countries (\textit{e.g.} Belgium and the Netherlands, or the Scandinavian countries), spatially clustered regions that span across countries (\textit{e.g.} in the UK) and spatially concentrated communities with very few overlap (in Germany). Nevertheless, the relatively larger communities break into smaller ones in all cases. This finding suggests that collaboration is relatively more likely to be repeated between geographically proximate locations while new collaboration is relatively more likely to bridge distant locations.

\subsection*{Gravity approaches of co-inventorship}

We further investigate the notion of different sizes of spatial communities in the complete \textit{versus} the persisted collaboration networks by applying two gravity approaches. In these gravity models, the unit of observation is the region-region link and we aim to identify the effect of dyadic covariates on collaboration and persisted collaboration. 

\vskip 0.1in
In the first gravity approach, that is often applied in network science\cite{liben2005geographic,lambiotte2008geographical,lengyel2015geographies}, we measure how distance decay, technological similarity and triadic closure of partnership increase or decrease the probability of collaboration and persisted collaboration across regions. Distance decay ($Distance$) influences network density\cite{liben2005geographic}, and thus yields spatial communities in and of itself \cite{expert2011uncovering}. Because technological similarity of regions and triadic closure of partnership influences the establishment and persistence of collaboration \cite{juhasz2017creation,maggioni2007space,ter2013dynamics}, we consider probabilities as a function of these dimensions as well. The overlap between the technological portfolios of regions is measured by cosine similarity ($Cosine$), taking the value of 0 in case of perfect mismatch and 1 in case of perfect overlap. Triadic closure is measured by Jaccard similarity ($Jaccard$), that takes the value of 0 if regions share no partners and 1 if all their partners are shared. The calculation of cosine and Jaccard similarity are described in detail in the Methods section.

\vskip 0.1in
We bin the distributions of region-region links into 20km intervals for $Distance$, 0.025 intervals for $Cosine$, and 0.01 intervals for $Jaccard$. Then, we calculate probability $P(Distance, Cosine, Jaccard)$ of collaboration and persisted collaboration for every group such that
\begin{equation}
$$P(Distance, Cosine, Jaccard) = \frac{\sum_{Distance, Cosine, Jaccard}L_{ij}}{\sum_{Distance, Cosine, Jaccard}N_i \times N_j}$$,
\end{equation} 
where $L_{ij}$ is the number of observed individual co-inventor connections between regions $i$ and $j$ in the corresponding network, whereas $N_i$ and $N_j$ refer to the number of inventors in these regions who authored at least one patent in the 2006-2010 period regardless whether they established a new collaboration or persisted an old one. 

\begin{figure}[!ht]
\centering
\includegraphics[width=\linewidth]{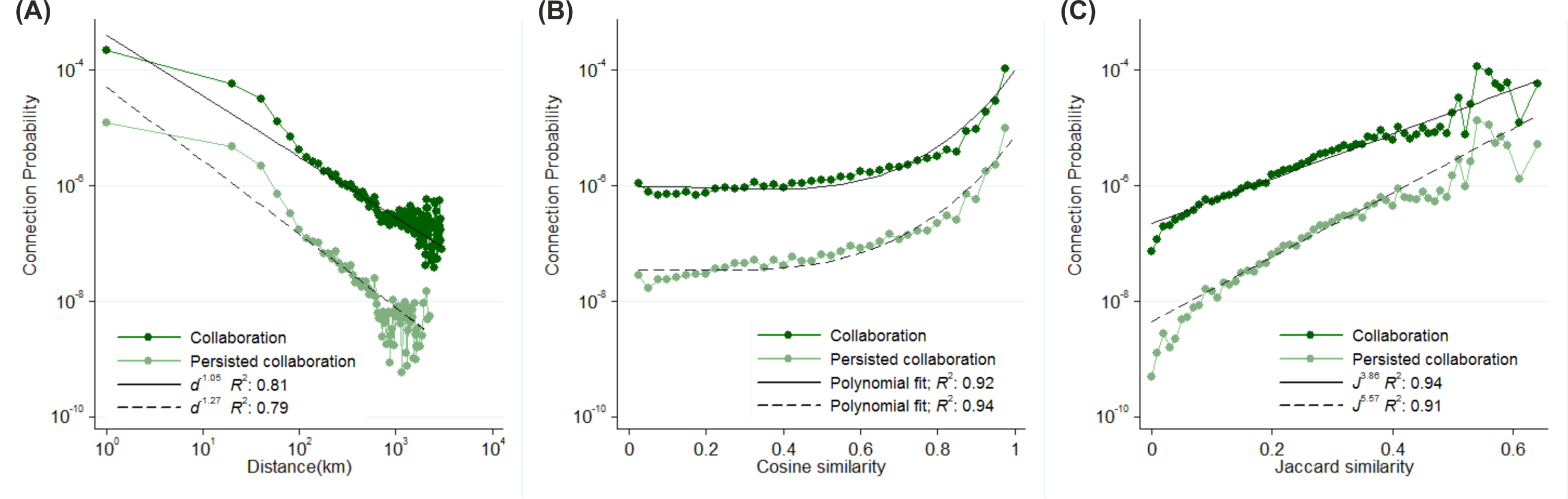}
\caption{The probability of collaboration and persisted collaboration as a function of region-to-region characteristics. \textbf{(A)} Distance decay is smooth for geographically proximate collaboration and persisted collaboration, and follows linear decay on log-log scale with the exponents -1.05 and -1.27 for distances larger than 100 km. \textbf{(B)} The overlap between technological profiles of regions, measured by cosine similarity, increases the probability of collaboration with a growing intensity as similarity rises. \textbf{(C)} The probability of collaboration grows linearly on a logarithmic scale as the share of common third partners, measured by Jaccard similarity, increases. The exponent is 3.86 for collaboration and 5.67 for persisted collaboration.}
\label{fig:Fig2}
\end{figure}

\vskip 0.1in
Figure \ref{fig:Fig2} depicts the probability of collaboration and persisted collaboration as a function of distance, technological similarity and shared third partners. Since we use an identical denominator for both networks, the smaller probability for persisted collaboration evidently follows from a lower density of individual connections. However, these differences are surprisingly stable across all three distributions. The exponent of distance decay in the persistent collaboration network is somewhat higher than in the complete collaboration network, however, these power laws can be fitted to the middle of the distance distribution only, and curves are remarkably similar for distances smaller than 100 km. A fractional polynomial fit ($1.75 Cosine^{3}+3.22 Cosine^{3}$) captures the relation between cosine similarity and $P(Cosine)$ 
very well (Figure 2B), and suggests that a wide overlap of technological profiles increases the probability of collaboration and persisted collaboration ties in a similar manner. Moreover, after a critically extensive technological similarity the probability increases exponentially.
A power-law can be fitted on the Jaccard distribution as well such that an increasing fraction of shared third partners induce $P(Jaccard)$ 
and the exponent is higher in the persisted collaboration (Figure 2C). However, the fit only works for the middle of the distribution and the effect of Jaccard similarity becomes remarkably alike across the networks when at least 40\% of third partners are shared. 

\vskip 0.1in
Because univariate gravity models do not reveal remarkable differences of newly created \textit{versus} persisted co-inventor collaborations, we apply a multivariate gravity approach often used in regional science and economic geography \cite{broekel2014modeling,maggioni2007space}. Distance decay correlates with triadic closure\cite{lambiotte2008geographical,ter2013dynamics}, and also with technological profiles of regions \cite{cassi2015research}; therefore, the multivariate approach is straightforward to disentangle different effects behind the spatial structure of collaboration and persisted collaboration networks.

\begin{figure}[!ht]
\centering
\includegraphics[width=120mm]{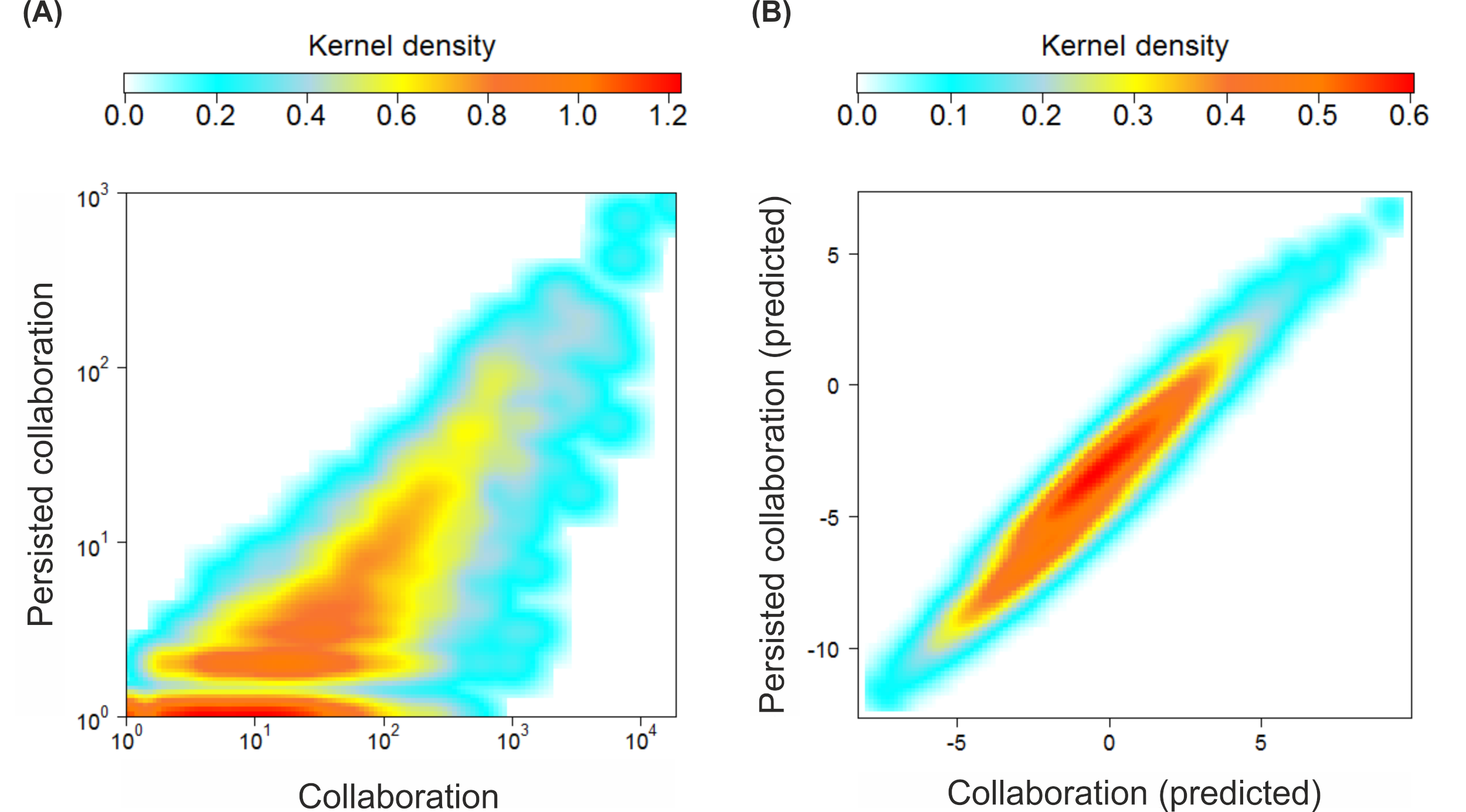}
\caption{\textbf{(A)} Tie strength correlation between total collaboration and persisted collaboration ties. \textbf{(B)} Marginals estimated from the gravity equation of the strength of collaboration and persisted collaboration across regions.}
\label{fig:Fig3}
\end{figure}

\vskip 0.1in
In multivariate gravity modeling, the strength of region-region links are estimated in a regression framework with dyadic covariates and characteristics of both regions involved in the dyad. We look for differences in the correlation values of co-variates between the strength of collaboration and persisted collaboration ties on the same set of region-region links, for which the co-variates are identical. Therefore, a necessary condition for the exercise is the variation of collaboration \textit{versus} persisted collaboration tie strengths. In Figure \ref{fig:Fig3}A we see a very strong correlation between collaboration and persisted collaboration. However, sufficient variance between these two values is also present. For example there is a relatively wide distribution of collaboration strength for those region-region links that have 10\textsuperscript{1} persisted ties: the minimum, mean and maximum values are around 10\textsuperscript{1}, 10\textsuperscript{2} and 10\textsuperscript{3}.

\vskip 0.1in
The strength of collaboration and persisted collaboration ties can be considered as count data of individual co-inventor edges, in which most region-region links account for zero individual connections. Therefore, we apply a Zero-Inflated Negative Binomial  (ZINB) regression, which consists of two parts\cite{greene1994accounting}. The first equation refers to the count process, in which we estimate the number of individual ties between regions by our three main variables. This equation is formulated as
\begin{equation}
$
$log(Y_{ij}=y_{ij})=\beta_{0}+\beta_{1}Proximity_{ij}+\beta_{2}Cosine_{ij}+\beta_{3}Jaccard_{ij}+\epsilon_{ij},$
$
\end{equation}
where $Y_{ij}$ 
is the strength of collaboration or persisted collaboration between regions $i$ and $j$, $Proximity_{ij}$ is the inverse of distance, while $Cosine_{ij}$ and $Jaccard_{ij}$ refer to variables explained above. For detailed description of the variables see the Methodology section. Besides transforming distance to proximity, which is important to have all correlations with identical sign, we standardize every variable so that coefficients sizes can be compared across varying scales.
The second equation is often referred to as regime selection, which is to deal with excessive zeros in the data and is formulated as 
\begin{equation}
$
$P(Y_{ij}=0)=\gamma_{0}+\gamma_{1}\theta_{ij}+\gamma_{2}log(Z_{i},Z_{j})+\epsilon_{ij},$
$
\end{equation}

where we estimate the probability that a connection may develop between two regions. In the equation $\theta_{ij}$ is a dummy variable that takes the value 1 if region \textit{i} and region \textit{j} are in the same country, and $Z_{i},Z_{j}$ are a collection of region-level control variables that are commonly used in similar estimations\cite{maggioni2007space}. The list and motivation of control variables and a more detailed description of ZINB regression models can be found in the Methods section.

\vskip 0.1in
In Figure \ref{fig:Fig3}B, we plot predicted values of collaboration ties against the predicted values of all collaborations and persisted collaboration ties calculated by estimating the coefficients of the ZINB regressions.
The skewed distribution suggests that the gravity estimation narrows down the variance we observe in Figure \ref{fig:Fig3}A. The two separate estimations, in which only the dependent variables and the Jaccard indexes differ, yield very similar predicted values of collaboration and persisted collaboration for region pairs. Therefore, if coefficients suggest different effects, one can argue for diverging mechanisms of establishing new inventor collaboration \textit{versus} persisting old collaborations. 

\begin{figure}[!ht]
\centering
\includegraphics[width=100mm]{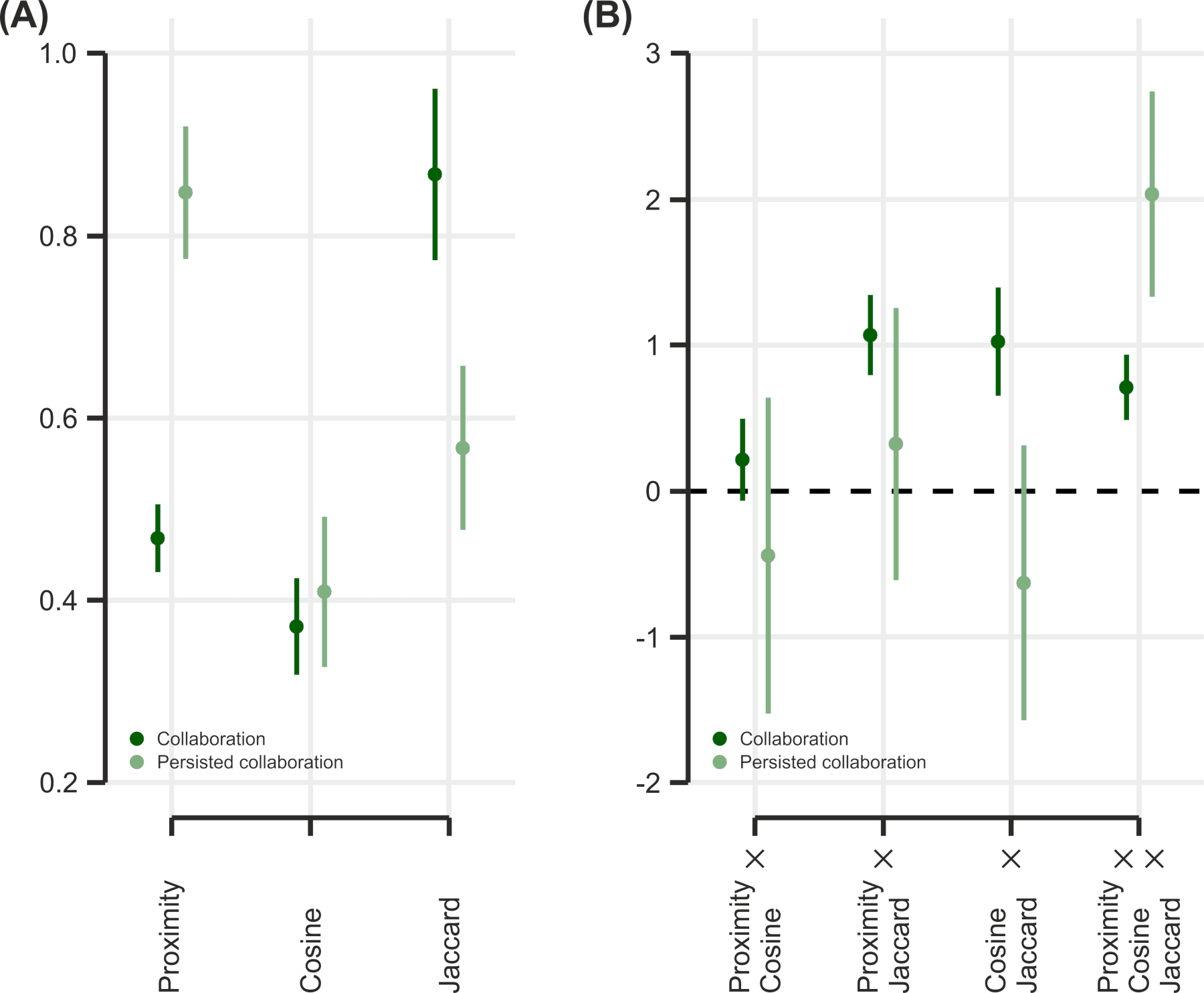}
\caption{Estimation results of the multivariate gravity equation. \textbf{(A)} Considering single effects only, we observe that the probability of collaboration is mostly increased by common third partners, while persistent ties gain probability if regions are geographically proximate. \textbf{(B)} Interaction effects reveal that persistent ties gain extra probability if regions are geographically proximate, technologically similar and share many partners at the same time.}
\label{fig:Fig4}
\end{figure}

\vskip 0.1in
The parameters estimated from Equation 3 are presented in Figure \ref{fig:Fig4}A with their 95\% confidence intervals. Because all three explanatory variables are standardized, coefficients show the expected change in the log number of  collaboration and persisted collaboration  in case there is one standard deviation change in the independent variable. The details of the parameter estimations are presented in Supporting Information 3. Results suggest that geographical proximity is more important to persist collaboration than what is observed in the collaboration network in general. Technological similarity does not differentiate collaboration from persisted collaboration across European regions. Interestingly, high shares of common third partners favour collaboration more but not so much the persistence of collaborations. These findings are in line with previous results of Ter wal\cite{ter2013dynamics}, who argues that common third partners decrease costs and uncertainties of new link formation, but once the collaboration is established, geographical proximity is more important in decreasing the costs of maintaining and thus further strengthening these relations. The importance of geographical proximity gives a hint already at why spatial communities of persisted networks are of smaller spatial scale than the ones in the general collaboration network.

\vskip 0.1in
However, the correlation between these variables\cite{lambiotte2008geographical,cassi2015research,juhasz2017creation} call for better understanding how their joint effect increases collaboration and persisted collaboration. Therefore, we include the interaction effect of our main variables and modify Equation 4 such that
\begin{equation}
\begin{aligned}
log(Y_{ij}=y_{ij})=
& \beta_{0}+
\beta_{1}Proximity_{ij}^{D}+
\beta_{2}Cosine_{ij}^{D}+
\beta_{3}Jaccard_{ij}^{D} + \\
& \beta_{4} (Proximity_{ij}^{D} \times Cosine_{ij}^{D}) + 
\beta_{5} (Proximity_{ij}^{D} \times Jaccard_{ij}^{D}) +
\beta_{6} (Cosine_{ij}^{D} \times Jaccard_{ij}^{D}) + \\
& \beta_{7} (Proximity_{ij}^{D}  \times Cosine_{ij}^{D} \times Jaccard_{ij}^{D})+\epsilon_{ij},
\end{aligned}
\end{equation}

where $Y_{ij}$ is the strength of collaboration or persisted collaboration, and the $D$ upper index refers to a dummy transformation of the variables after which the variable gets the value of 1 in case it is higher or equal to the mean of its distribution and 0 otherwise. This transformation is useful since the sign of those interaction coefficients that significantly differ from zero can be interpreted to influence collaboration positively or negatively, while the interpretations of the interactions of the continuous variables would not be straightforward.

\vskip 0.1in
In Figure \ref{fig:Fig4}B, we illustrate the coefficients and 95\% confidence interval of interaction terms obtained from estimating the ZINB regression of Equations 5. Details of the estimations are presented in SI 4. Interaction coefficients of collaboration ties are positive and significant in all cases when the Jaccard index is involved. This means that sharing third partners increases the effect of both geographical proximity and technological similarity. More importantly, we find that persisted collaboration is increased by interaction terms only if all three variables are included in the joint effect. This last finding suggests that the persistence of cross-regional inventor collaboration in Europe gains extra likelihood when regions are geographically proximate, are similar in their technological profile and share most of their third partners. Consequently, the communities of persisted collaboration are of smaller spatial scale not only because the effect of geographical proximity is at work, but also because this is coupled with technological similarity and shared third partners.


\vskip 0.1in

\section*{Discussion}

\vskip 0.1in

The focus of this study has been on the nature of co-patenting  and the geography of persisted collaboration across European regions. We found that only a fraction of patent collaborations across regions are persisted, and they are clustered more intensively compared to the complete network. Our results show that new collaborations in inter-region co-patenting network emerge mainly through triadic closure, while geographical proximity becomes the most influential factor for maintaining co-patenting. In addition to that, the combination of technological similarity and shared partners offer a premium for the likelihood to maintain collaboration, but only when geographical proximity is present as an enabler. All together these findings suggest that there is an intensive search process for new partners in the inter-region collaboration network under study, where the main strategy for decreasing the uncertainty of benefits from collaboration is through partners of partners becoming partners. However the repeated cost of lasting collaborations shifts emphasis to geographical proximity and the added benefit of multidimensional proximity among partners.

\vskip 0.1in
The findings of this paper bear consequences for innovation policy in the EU. The declared objective of establishing a European Research Area was to decrease the fragmentation of the European research activities, that are still self-organized into national innovation systems. However, those inter-regional collaborations, that are deemed worthy of maintaining are fragmented at a smaller spatial scale of regional innovation systems, that also follow national borders. This poses challenges for both establishing cross-border regional innovation systems and for developing lasting cooperations between more distant places. While the spatial clustering of innovative activities offers positive externalities \cite{morescalchi2015evolution}, fostering a diverse and open network of inventors in general helps in exploring opportunities \cite{crescenzi2016inventors}, and in avoiding lock-in. However, our evidence suggests that lasting cooperations are strongly bounded by multidimensional proximity, and are likely to revert to spatial clustering. European innovation policy could address these issues by specifically targeting persisted collaborations across national borders and larger distances.

\vskip 0.1in

\section*{Methods}

\begin{figure}[!hb]
\centering
\includegraphics[width=\linewidth]{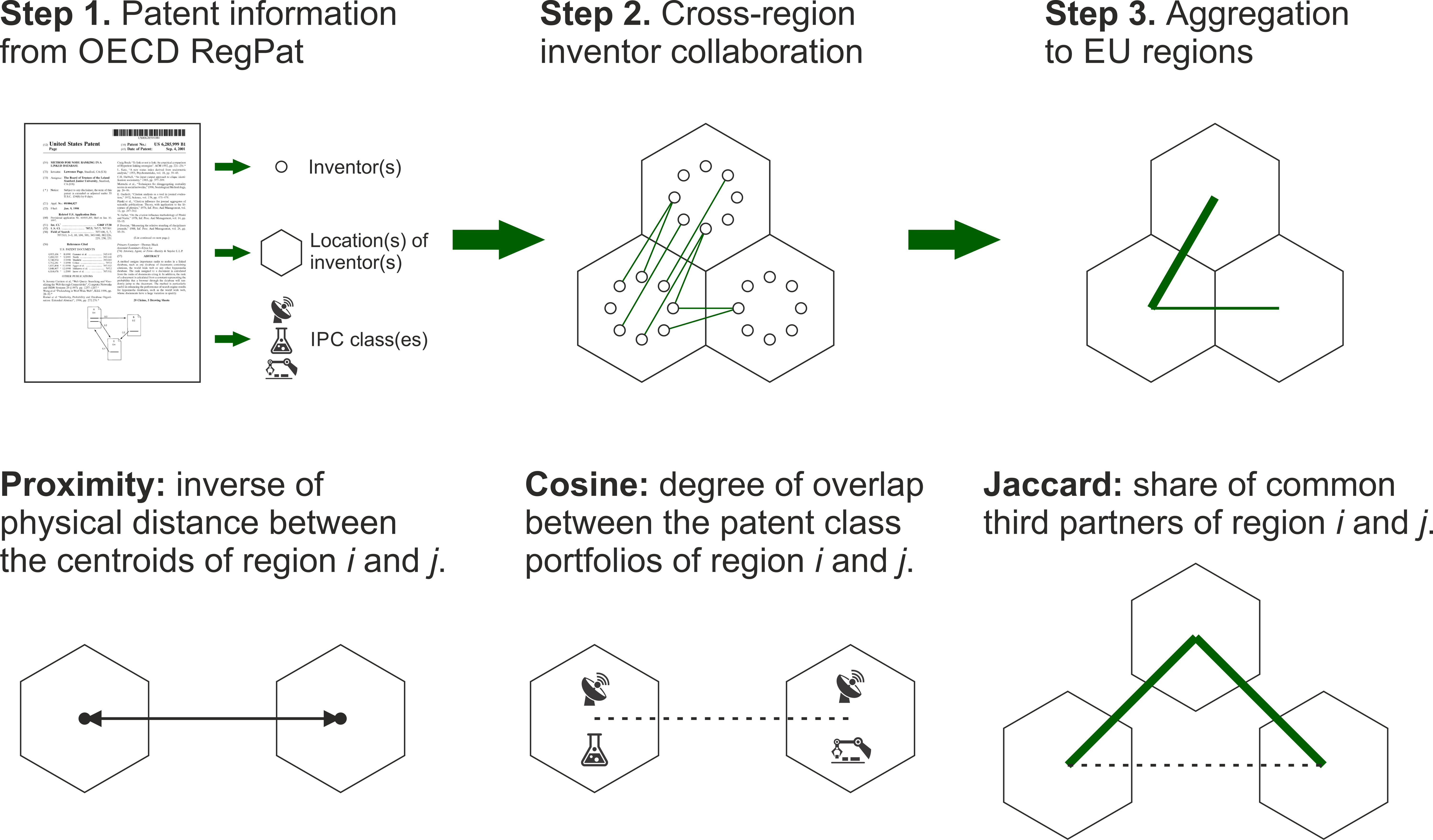}
\caption{ Data management and variable creation.}
\label{fig:Fig5}
\end{figure}

\vskip 0.1in
Figure \ref{fig:Fig5} offers an overview of the data management process and the key explanatory variables. We used the publicly available OECD REGPAT database for all patent classifications. For every patent the database includes a unique ID and location information at NUTS3 level of the inventor(s) and the IPC classification(s). From these patent informations we derived the cross regional inventor collaboration network. The tie weights are proportional to the number of inventor collaborations between regions in the 2006 to 2010 period. We consider an inventor-inventor tie persisted when it has occurred at least once in the 1991-2005 period, priory our investigation. All data is publicly available upon request from OECD. The region-region edgelists, from which our results can be reproduced are available on this \href{https://drive.google.com/drive/folders/1ReJjffQVbmkB5PIv2n1qnivmX_oGJ6jr?usp=sharing}{link}.

\vskip 0.1in
The \textit{Distance} variable denotes physical distance measured in kilometers between the centroids of each region. Geographical \textit{Proximity} between regions $i$ and $j$ was obtained by using the inverse of \textit{Distance} calculated as $maximum(Distance_{ij})-Distance_{ij}$.

\vskip 0.1in
A common way of quantifying technological similarity between two regions $i$ and $j$ is to calculate the cosine similarity of the patent portfolio vectors $V_{i}$ and $V_{j}$. The variable \textit{Cosine} is defined as:
\begin{equation}
$
$Cosine_{ij}=\frac{V_{i}\cdot V_{j}}{|\sum_{i=1}^{M}{V_{i}^{2}}||\sum_{j=1}^{M}{V_{j}^{2}}|},$
$
\end{equation}
where the numerator is the inner product of the regional patent portfolios. These portfolios  contain the number of patents for each 4-digit IPC class in the region. The denominator is the product of the Euclidean length of each patent portfolio vector \cite{schutze2008introduction}. The theoretical range of cosine similarity is the $[-1,1]$ interval, however as patent numbers cannot take on a negative value our variable ranges from zero to one, where zero is the case of perfectly unrelated portfolios and one represents complete similarity.

\vskip 0.1in
To calculate shared partners in two regions' collaboration network we used the Jaccard index, that measures the overlap between finite sample sets, and is defined as the cardinality of the intersection divided by the cardinality of the union\cite{leydesdorff2008normalization}:

\begin{equation}
$
$Jaccard_{ij}=\frac{|A_{i}\cap A_{j}|}{|A_{i}\cup A_{j}|},$
$
\end{equation}
where $A_{i}$ and $A_{j}$ refer to the underlying collaboration matrices of regions $i$ and $j$. The \textit{Jaccard} variable ranges from zero to one, where higher values indicate a higher share of common third partners. We also calculated Dice similarity and inverse log-weighted similarity that are alternatives for Jaccard; however, we found a very strong correlation between these indices (above 0.9 in all cases). Therefore we opted for the Jaccard index because it is the most intuitively interpretable.

\vskip 0.1in
The comparison of the three main continuous explanatory variables may lead to misleading inferences. One unit change in Cosine similarity is hardly comparable with one unit change in Jaccard index or one kilometer change in geographic proximity and vice-versa. Standardizing continuous variables makes it easier to compare coefficients of the estimations on the same scale. Therefore, the three main explanatory variables have been rescaled to have a mean of zero with one standard deviation. We used the following standardization function  $z(x)=\frac{(x-\overline{x})}{\sigma_{x}}$,where $\overline{x}$ is the mean of $x$, and $\sigma_{x}$ is the standard deviation of $x$.

\vskip 0.1in
Co-inventor collaborations between regions can be considered as a count process. Because there are an excessive number region-region pairs with zero collaboration ties, we have to deal with the cases of missing collaborations independently as creating zero collaboration would be a distinct process from creating non-zero collaboration. In the Zero Inflated Negative Binomial regression technique, we assume that missing collaboration in a region-region pair is a result of a conscious decision of inventors. In this model, $P_{ij}$ is the probability that the value of observation \textit{ij} is zero and the count process is governed with probability (1-$P_{i}$) by a negative binomial distribution with mean $\lambda_{i}$. Therefore the response $Y_{ij}$ follows the distribution of:

\begin{equation}
$
$Y_{i}$ = 
$\begin{cases}
    0,	& \text{with probability }  P_{ij}+(1-P_{ij})e^{-\lambda_{ij}}  \\
    y_{ij},  & \text{with probability }  (1-P_{ij})\frac{e^{-\lambda_{ij}}{\lambda_{ij}^{y_{ij}}}}{y_{ij}} , y_{ij}=1,2...
\end{cases}$
$
\end{equation}

\vskip 0.1in
The ZINB model is therefore a combination of a probability function and a negative binomial distribution where $\lambda_{ij}$ is the mean and $k$ is the overdispersion parameter. Following Greene\cite{greene1994accounting} the model can be written as:

\begin{equation}
$
$P(Y_{ij}=0)=P_{ij}+(1-P_{ij})(1+k\lambda_{ij})^{-1/k}$
$
\end{equation}
\begin{equation}
$
$P(Y_{ij}=y_{ij})=(1-P_{ij})\frac{\Gamma(y_{ij}+1/k)(k\lambda_{ij})^{y_{ij}}}{\Gamma(y_{ij}+1)\Gamma(1/k)(1+k\lambda_{ij})^{y_{ij}1/k}}, y_{ij}=1,2...$
$
\end{equation}

\vskip 0.1in 
In this paper, the inflation part of the regression consisted of variables that usually represent magnitude in a multivariate gravity approach, such as the number of inventors, population density, the log of gross value added and the aggregate number of co-inventor collaboration, plus an additional dummy variable for within country cooperations. We inserted the explanatory variables into the count process part of the model.

\pagebreak
\bibliography{references}

\begin{thebibliography}{10}
\expandafter\ifx\csname url\endcsname\relax
  \def\url#1{\texttt{#1}}\fi
\expandafter\ifx\csname urlprefix\endcsname\relax\def\urlprefix{URL }\fi
\expandafter\ifx\csname doiprefix\endcsname\relax\def\doiprefix{DOI }\fi
\providecommand{\bibinfo}[2]{#2}
\providecommand{\eprint}[2][]{\url{#2}}

\bibitem{audretsch1996r}
\bibinfo{author}{Audretsch, D.~B.} \& \bibinfo{author}{Feldman, M.~P.}
\newblock \bibinfo{journal}{\bibinfo{title}{R\&d spillovers and the geography
  of innovation and production}}.
\newblock {\emph{\JournalTitle{The American economic review}}}
  \textbf{\bibinfo{volume}{86}}, \bibinfo{pages}{630--640}
  (\bibinfo{year}{1996}).

\bibitem{bettencourt2007growth}
\bibinfo{author}{Bettencourt, L.~M.}, \bibinfo{author}{Lobo, J.},
  \bibinfo{author}{Helbing, D.}, \bibinfo{author}{K{\"u}hnert, C.} \&
  \bibinfo{author}{West, G.~B.}
\newblock \bibinfo{journal}{\bibinfo{title}{Growth, innovation, scaling, and
  the pace of life in cities}}.
\newblock {\emph{\JournalTitle{Proceedings of the national academy of
  sciences}}} \textbf{\bibinfo{volume}{104}}, \bibinfo{pages}{7301--7306}
  (\bibinfo{year}{2007}).

\bibitem{anselin1997local}
\bibinfo{author}{Anselin, L.}, \bibinfo{author}{Varga, A.} \&
  \bibinfo{author}{Acs, Z.}
\newblock \bibinfo{journal}{\bibinfo{title}{Local geographic spillovers between
  university research and high technology innovations}}.
\newblock {\emph{\JournalTitle{Journal of urban economics}}}
  \textbf{\bibinfo{volume}{42}}, \bibinfo{pages}{422--448}
  (\bibinfo{year}{1997}).

\bibitem{balland2017geography}
\bibinfo{author}{Balland, P.-A.} \& \bibinfo{author}{Rigby, D.}
\newblock \bibinfo{journal}{\bibinfo{title}{The geography of complex
  knowledge}}.
\newblock {\emph{\JournalTitle{Economic Geography}}}
  \textbf{\bibinfo{volume}{93}}, \bibinfo{pages}{1--23} (\bibinfo{year}{2017}).

\bibitem{jaffe1993geographic}
\bibinfo{author}{Jaffe, A.~B.}, \bibinfo{author}{Trajtenberg, M.} \&
  \bibinfo{author}{Henderson, R.}
\newblock \bibinfo{journal}{\bibinfo{title}{Geographic localization of
  knowledge spillovers as evidenced by patent citations}}.
\newblock {\emph{\JournalTitle{the Quarterly journal of Economics}}}
  \textbf{\bibinfo{volume}{108}}, \bibinfo{pages}{577--598}
  (\bibinfo{year}{1993}).

\bibitem{storper2004buzz}
\bibinfo{author}{Storper, M.} \& \bibinfo{author}{Venables, A.~J.}
\newblock \bibinfo{journal}{\bibinfo{title}{Buzz: face-to-face contact and the
  urban economy}}.
\newblock {\emph{\JournalTitle{Journal of economic geography}}}
  \textbf{\bibinfo{volume}{4}}, \bibinfo{pages}{351--370}
  (\bibinfo{year}{2004}).

\bibitem{katz1985network}
\bibinfo{author}{Katz, M.~L.} \& \bibinfo{author}{Shapiro, C.}
\newblock \bibinfo{journal}{\bibinfo{title}{Network externalities, competition,
  and compatibility}}.
\newblock {\emph{\JournalTitle{The American economic review}}}
  \textbf{\bibinfo{volume}{75}}, \bibinfo{pages}{424--440}
  (\bibinfo{year}{1985}).

\bibitem{acemoglu2016innovation}
\bibinfo{author}{Acemoglu, D.}, \bibinfo{author}{Akcigit, U.} \&
  \bibinfo{author}{Kerr, W.~R.}
\newblock \bibinfo{journal}{\bibinfo{title}{Innovation network}}.
\newblock {\emph{\JournalTitle{Proceedings of the National Academy of
  Sciences}}} \textbf{\bibinfo{volume}{113}}, \bibinfo{pages}{11483--11488}
  (\bibinfo{year}{2016}).

\bibitem{breschi2009mobility}
\bibinfo{author}{Breschi, S.} \& \bibinfo{author}{Lissoni, F.}
\newblock \bibinfo{journal}{\bibinfo{title}{Mobility of skilled workers and
  co-invention networks: an anatomy of localized knowledge flows}}.
\newblock {\emph{\JournalTitle{Journal of economic geography}}}
  \textbf{\bibinfo{volume}{9}}, \bibinfo{pages}{439--468}
  (\bibinfo{year}{2009}).

\bibitem{sorenson2006complexity}
\bibinfo{author}{Sorenson, O.}, \bibinfo{author}{Rivkin, J.~W.} \&
  \bibinfo{author}{Fleming, L.}
\newblock \bibinfo{journal}{\bibinfo{title}{Complexity, networks and knowledge
  flow}}.
\newblock {\emph{\JournalTitle{Research policy}}}
  \textbf{\bibinfo{volume}{35}}, \bibinfo{pages}{994--1017}
  (\bibinfo{year}{2006}).

\bibitem{cantner2017coevolution}
\bibinfo{author}{Cantner, U.}, \bibinfo{author}{Hinzmann, S.} \&
  \bibinfo{author}{Wolf, T.}
\newblock \bibinfo{title}{The coevolution of innovative ties, proximity, and
  competencies: Toward a dynamic approach to innovation cooperation}.
\newblock In \emph{\bibinfo{booktitle}{Knowledge and Networks}},
  \bibinfo{pages}{337--372} (\bibinfo{publisher}{Springer},
  \bibinfo{year}{2017}).

\bibitem{cassi2015research}
\bibinfo{author}{Cassi, L.} \& \bibinfo{author}{Plunket, A.}
\newblock \bibinfo{journal}{\bibinfo{title}{Research collaboration in
  co-inventor networks: combining closure, bridging and proximities}}.
\newblock {\emph{\JournalTitle{Regional Studies}}}
  \textbf{\bibinfo{volume}{49}}, \bibinfo{pages}{936--954}
  (\bibinfo{year}{2015}).

\bibitem{balland2012proximity}
\bibinfo{author}{Balland, P.-A.}
\newblock \bibinfo{journal}{\bibinfo{title}{Proximity and the evolution of
  collaboration networks: evidence from research and development projects
  within the global navigation satellite system (gnss) industry}}.
\newblock {\emph{\JournalTitle{Regional Studies}}}
  \textbf{\bibinfo{volume}{46}}, \bibinfo{pages}{741--756}
  (\bibinfo{year}{2012}).

\bibitem{brenner2013introduction}
\bibinfo{author}{Brenner, T.}, \bibinfo{author}{Cantner, U.} \&
  \bibinfo{author}{Graf, H.}
\newblock \bibinfo{journal}{\bibinfo{title}{Introduction: Structure and
  dynamics of innovation networks}}.
\newblock {\emph{\JournalTitle{Regional Studies}}}
  \textbf{\bibinfo{volume}{47}}, \bibinfo{pages}{647--650}
  (\bibinfo{year}{2013}).

\bibitem{fleming2007evolution}
\bibinfo{author}{Fleming, L.} \& \bibinfo{author}{Frenken, K.}
\newblock \bibinfo{journal}{\bibinfo{title}{The evolution of inventor networks
  in the silicon valley and boston regions}}.
\newblock {\emph{\JournalTitle{Advances in Complex Systems}}}
  \textbf{\bibinfo{volume}{10}}, \bibinfo{pages}{53--71}
  (\bibinfo{year}{2007}).

\bibitem{fritsch2004innovation}
\bibinfo{author}{Fritsch, M.} \& \bibinfo{author}{Franke, G.}
\newblock \bibinfo{journal}{\bibinfo{title}{Innovation, regional knowledge
  spillovers and r\&d cooperation}}.
\newblock {\emph{\JournalTitle{Research policy}}}
  \textbf{\bibinfo{volume}{33}}, \bibinfo{pages}{245--255}
  (\bibinfo{year}{2004}).

\bibitem{maggioni2007space}
\bibinfo{author}{Maggioni, M.~A.}, \bibinfo{author}{Nosvelli, M.} \&
  \bibinfo{author}{Uberti, T.~E.}
\newblock \bibinfo{journal}{\bibinfo{title}{Space versus networks in the
  geography of innovation: A european analysis}}.
\newblock {\emph{\JournalTitle{Papers in Regional Science}}}
  \textbf{\bibinfo{volume}{86}}, \bibinfo{pages}{471--493}
  (\bibinfo{year}{2007}).

\bibitem{ter2013dynamics}
\bibinfo{author}{Ter~Wal, A.~L.}
\newblock \bibinfo{journal}{\bibinfo{title}{The dynamics of the inventor
  network in german biotechnology: geographic proximity versus triadic
  closure}}.
\newblock {\emph{\JournalTitle{Journal of Economic Geography}}}
  \textbf{\bibinfo{volume}{14}}, \bibinfo{pages}{589--620}
  (\bibinfo{year}{2013}).

\bibitem{bathelt2004clusters}
\bibinfo{author}{Bathelt, H.}, \bibinfo{author}{Malmberg, A.} \&
  \bibinfo{author}{Maskell, P.}
\newblock \bibinfo{journal}{\bibinfo{title}{Clusters and knowledge: local buzz,
  global pipelines and the process of knowledge creation}}.
\newblock {\emph{\JournalTitle{Progress in human geography}}}
  \textbf{\bibinfo{volume}{28}}, \bibinfo{pages}{31--56}
  (\bibinfo{year}{2004}).

\bibitem{gluckler2007economic}
\bibinfo{author}{Gl{\"u}ckler, J.}
\newblock \bibinfo{journal}{\bibinfo{title}{Economic geography and the
  evolution of networks}}.
\newblock {\emph{\JournalTitle{Journal of Economic Geography}}}
  \textbf{\bibinfo{volume}{7}}, \bibinfo{pages}{619--634}
  (\bibinfo{year}{2007}).

\bibitem{grabher1993weakness}
\bibinfo{author}{Grabher, G.}
\newblock \bibinfo{journal}{\bibinfo{title}{The weakness of strong ties; the
  lock-in of regional development in ruhr area}}.
\newblock {\emph{\JournalTitle{The embedded firm; on the socioeconomics of
  industrial networks}}} \bibinfo{pages}{255--277} (\bibinfo{year}{1993}).

\bibitem{boschma2010spatial}
\bibinfo{author}{Boschma, R.} \& \bibinfo{author}{Frenken, K.}
\newblock \bibinfo{journal}{\bibinfo{title}{The spatial evolution of innovation
  networks. a proximity perspective}}.
\newblock {\emph{\JournalTitle{The handbook of evolutionary economic
  geography}}} \bibinfo{pages}{120--135} (\bibinfo{year}{2010}).

\bibitem{uzzi1997social}
\bibinfo{author}{Uzzi, B.}
\newblock \bibinfo{journal}{\bibinfo{title}{Social structure and competition in
  interfirm networks: The paradox of embeddedness}}.
\newblock {\emph{\JournalTitle{Administrative science quarterly}}}
  \bibinfo{pages}{35--67} (\bibinfo{year}{1997}).

\bibitem{juhasz2017creation}
\bibinfo{author}{Juh{\'a}sz, S.} \& \bibinfo{author}{Lengyel, B.}
\newblock \bibinfo{journal}{\bibinfo{title}{Creation and persistence of ties in
  cluster knowledge networks}}.
\newblock {\emph{\JournalTitle{Journal of Economic Geography}}}
  (\bibinfo{year}{2017}).

\bibitem{dahlander2013ties}
\bibinfo{author}{Dahlander, L.} \& \bibinfo{author}{McFarland, D.~A.}
\newblock \bibinfo{journal}{\bibinfo{title}{Ties that last: Tie formation and
  persistence in research collaborations over time}}.
\newblock {\emph{\JournalTitle{Administrative science quarterly}}}
  \textbf{\bibinfo{volume}{58}}, \bibinfo{pages}{69--110}
  (\bibinfo{year}{2013}).

\bibitem{rivera2010dynamics}
\bibinfo{author}{Rivera, M.~T.}, \bibinfo{author}{Soderstrom, S.~B.} \&
  \bibinfo{author}{Uzzi, B.}
\newblock \bibinfo{journal}{\bibinfo{title}{Dynamics of dyads in social
  networks: Assortative, relational, and proximity mechanisms}}.
\newblock {\emph{\JournalTitle{annual Review of Sociology}}}
  \textbf{\bibinfo{volume}{36}}, \bibinfo{pages}{91--115}
  (\bibinfo{year}{2010}).

\bibitem{chessa2013europe}
\bibinfo{author}{Chessa, A.} \emph{et~al.}
\newblock \bibinfo{journal}{\bibinfo{title}{Is europe evolving toward an
  integrated research area?}}
\newblock {\emph{\JournalTitle{Science}}} \textbf{\bibinfo{volume}{339}},
  \bibinfo{pages}{650--651} (\bibinfo{year}{2013}).

\bibitem{arrieta2017quantifying}
\bibinfo{author}{Arrieta, O. A.~D.}, \bibinfo{author}{Pammolli, F.} \&
  \bibinfo{author}{Petersen, A.~M.}
\newblock \bibinfo{journal}{\bibinfo{title}{Quantifying the negative impact of
  brain drain on the integration of european science}}.
\newblock {\emph{\JournalTitle{Science Advances}}}
  \textbf{\bibinfo{volume}{3}}, \bibinfo{pages}{e1602232}
  (\bibinfo{year}{2017}).

\bibitem{lengyel2016international}
\bibinfo{author}{Lengyel, B.} \& \bibinfo{author}{Lesk{\'o}, M.}
\newblock \bibinfo{journal}{\bibinfo{title}{International collaboration and
  spatial dynamics of us patenting in central and eastern europe 1981-2010}}.
\newblock {\emph{\JournalTitle{PLoS One}}} \textbf{\bibinfo{volume}{11}}
  (\bibinfo{year}{2016}).

\bibitem{blondel2008fast}
\bibinfo{author}{Blondel, V.~D.}, \bibinfo{author}{Guillaume, J.-L.},
  \bibinfo{author}{Lambiotte, R.} \& \bibinfo{author}{Lefebvre, E.}
\newblock \bibinfo{journal}{\bibinfo{title}{Fast unfolding of communities in
  large networks}}.
\newblock {\emph{\JournalTitle{Journal of statistical mechanics: theory and
  experiment}}} \textbf{\bibinfo{volume}{2008}}, \bibinfo{pages}{P10008}
  (\bibinfo{year}{2008}).

\bibitem{liben2005geographic}
\bibinfo{author}{Liben-Nowell, D.}, \bibinfo{author}{Novak, J.},
  \bibinfo{author}{Kumar, R.}, \bibinfo{author}{Raghavan, P.} \&
  \bibinfo{author}{Tomkins, A.}
\newblock \bibinfo{journal}{\bibinfo{title}{Geographic routing in social
  networks}}.
\newblock {\emph{\JournalTitle{Proceedings of the National Academy of Sciences
  of the United States of America}}} \textbf{\bibinfo{volume}{102}},
  \bibinfo{pages}{11623--11628} (\bibinfo{year}{2005}).

\bibitem{lambiotte2008geographical}
\bibinfo{author}{Lambiotte, R.} \emph{et~al.}
\newblock \bibinfo{journal}{\bibinfo{title}{Geographical dispersal of mobile
  communication networks}}.
\newblock {\emph{\JournalTitle{Physica A: Statistical Mechanics and its
  Applications}}} \textbf{\bibinfo{volume}{387}}, \bibinfo{pages}{5317--5325}
  (\bibinfo{year}{2008}).

\bibitem{lengyel2015geographies}
\bibinfo{author}{Lengyel, B.}, \bibinfo{author}{Varga, A.},
  \bibinfo{author}{S{\'a}gv{\'a}ri, B.}, \bibinfo{author}{Jakobi, {\'A}.} \&
  \bibinfo{author}{Kert{\'e}sz, J.}
\newblock \bibinfo{journal}{\bibinfo{title}{Geographies of an online social
  network}}.
\newblock {\emph{\JournalTitle{PloS one}}} \textbf{\bibinfo{volume}{10}},
  \bibinfo{pages}{e0137248} (\bibinfo{year}{2015}).

\bibitem{broekel2014modeling}
\bibinfo{author}{Broekel, T.}, \bibinfo{author}{Balland, P.-A.},
  \bibinfo{author}{Burger, M.} \& \bibinfo{author}{van Oort, F.}
\newblock \bibinfo{journal}{\bibinfo{title}{Modeling knowledge networks in
  economic geography: a discussion of four methods}}.
\newblock {\emph{\JournalTitle{The annals of regional science}}}
  \textbf{\bibinfo{volume}{53}}, \bibinfo{pages}{423--452}
  (\bibinfo{year}{2014}).

\bibitem{newman2006modularity}
\bibinfo{author}{Newman, M.~E.}
\newblock \bibinfo{journal}{\bibinfo{title}{Modularity and community structure
  in networks}}.
\newblock {\emph{\JournalTitle{Proceedings of the national academy of
  sciences}}} \textbf{\bibinfo{volume}{103}}, \bibinfo{pages}{8577--8582}
  (\bibinfo{year}{2006}).

\bibitem{sah2017unraveling}
\bibinfo{author}{Sah, P.}, \bibinfo{author}{Leu, S.~T.},
  \bibinfo{author}{Cross, P.~C.}, \bibinfo{author}{Hudson, P.~J.} \&
  \bibinfo{author}{Bansal, S.}
\newblock \bibinfo{journal}{\bibinfo{title}{Unraveling the disease consequences
  and mechanisms of modular structure in animal social networks}}.
\newblock {\emph{\JournalTitle{Proceedings of the National Academy of
  Sciences}}} \bibinfo{pages}{201613616} (\bibinfo{year}{2017}).

\bibitem{sobolevsky2013delineating}
\bibinfo{author}{Sobolevsky, S.} \emph{et~al.}
\newblock \bibinfo{journal}{\bibinfo{title}{Delineating geographical regions
  with networks of human interactions in an extensive set of countries}}.
\newblock {\emph{\JournalTitle{PloS one}}} \textbf{\bibinfo{volume}{8}},
  \bibinfo{pages}{e81707} (\bibinfo{year}{2013}).

\bibitem{jun2017meet}
\bibinfo{author}{Jun, B.}, \bibinfo{author}{Pinheiro, F.},
  \bibinfo{author}{Buchmann, T.}, \bibinfo{author}{Yi, S.-k.} \&
  \bibinfo{author}{Hidalgo, C.}
\newblock \bibinfo{journal}{\bibinfo{title}{Meet me in the middle: The
  reunification of germany's research network}}.
\newblock {\emph{\JournalTitle{arXiv preprint arXiv:1704.08426}}}
  (\bibinfo{year}{2017}).

\bibitem{expert2011uncovering}
\bibinfo{author}{Expert, P.}, \bibinfo{author}{Evans, T.~S.},
  \bibinfo{author}{Blondel, V.~D.} \& \bibinfo{author}{Lambiotte, R.}
\newblock \bibinfo{journal}{\bibinfo{title}{Uncovering space-independent
  communities in spatial networks}}.
\newblock {\emph{\JournalTitle{Proceedings of the National Academy of
  Sciences}}} \textbf{\bibinfo{volume}{108}}, \bibinfo{pages}{7663--7668}
  (\bibinfo{year}{2011}).

\bibitem{greene1994accounting}
\bibinfo{author}{Greene, W.~H.}
\newblock \bibinfo{title}{Accounting for excess zeros and sample selection in
  poisson and negative binomial regression models}.
\newblock \bibinfo{type}{Tech. Rep.} (\bibinfo{year}{1994}).

\bibitem{morescalchi2015evolution}
\bibinfo{author}{Morescalchi, A.}, \bibinfo{author}{Pammolli, F.},
  \bibinfo{author}{Penner, O.}, \bibinfo{author}{Petersen, A.~M.} \&
  \bibinfo{author}{Riccaboni, M.}
\newblock \bibinfo{journal}{\bibinfo{title}{The evolution of networks of
  innovators within and across borders: Evidence from patent data}}.
\newblock {\emph{\JournalTitle{Research Policy}}}
  \textbf{\bibinfo{volume}{44}}, \bibinfo{pages}{651--668}
  (\bibinfo{year}{2015}).

\bibitem{crescenzi2016inventors}
\bibinfo{author}{Crescenzi, R.}, \bibinfo{author}{Nathan, M.} \&
  \bibinfo{author}{Rodr{\'\i}guez-Pose, A.}
\newblock \bibinfo{journal}{\bibinfo{title}{Do inventors talk to strangers? on
  proximity and collaborative knowledge creation}}.
\newblock {\emph{\JournalTitle{Research Policy}}}
  \textbf{\bibinfo{volume}{45}}, \bibinfo{pages}{177--194}
  (\bibinfo{year}{2016}).

\bibitem{schutze2008introduction}
\bibinfo{author}{Sch{\"u}tze, H.}, \bibinfo{author}{Manning, C.~D.} \&
  \bibinfo{author}{Raghavan, P.}
\newblock \emph{\bibinfo{title}{Introduction to information retrieval}},
  vol.~\bibinfo{volume}{39} (\bibinfo{publisher}{Cambridge University Press},
  \bibinfo{year}{2008}).

\bibitem{leydesdorff2008normalization}
\bibinfo{author}{Leydesdorff, L.}
\newblock \bibinfo{journal}{\bibinfo{title}{On the normalization and
  visualization of author co-citation data: Salton's cosine versus the jaccard
  index}}.
\newblock {\emph{\JournalTitle{Journal of the American Society for Information
  Science and Technology}}} \textbf{\bibinfo{volume}{59}},
  \bibinfo{pages}{77--85} (\bibinfo{year}{2008}).

\end{thebibliography}

\section*{Acknowledgements}

The authors acknowledge comments received at the MIT Media Lab workshop on "Innovation, Complexity and Economic Geography".

\section*{Author contributions statement}

All authors designed the research. G.T., S.J. and Z. E. cleaned and managed the data, G.T and B.L. made the analyses, G.T, Z.E and B.L prepared the illustrations, G.T, Z.E and B.L wrote the paper. All authors reviewed the manuscript. 

\section*{Additional information}

\textbf{Competing financial interests}: The authors declare no competing financial interest. 

\pagebreak
\section*{Supporting information}
\beginsupplement

\begin{table}[ht]
\caption {Rand index between alternative community structures of the full co-inventor network} \label{tab:SI1}
\centering
\begin{tabular}{l|llll}
& 2 & 3 & 4 & 5 \\
\hline
1 & 0.703 & 0.722 & 0.737 & 0.741 \\
2 &  & 0.658 & 0.671 & 0.673 \\
3 & & & 0.685 & 0.687 \\
4 & & & & 0.702 \\
\end{tabular}
\end{table}

\begin{table}[ht]
\caption {Rand index between alternative community structures of the persistent co-inventor network} \label{tab:SI2}
\centering
\begin{tabular}{l|llll}
& 2 & 3 & 4 & 5 \\
\hline
1 & 0.811 & 0.808 & 0.769 & 0.779 \\
2 &  & 0.724 & 0.693 & 0.702 \\
3 & & & 0.692 & 0.700 \\
4 & & & & 0.674 \\
\end{tabular}
\end{table}

\begin{table}[ht]
\caption {Multivariate gravity models, zero-inflated negative binomial regression} \label{tab:SI3} 
\centering
\def\sym#1{\ifmmode^{#1}\else\(^{#1}\)\fi}
\begin{tabular}{l*{2}{c}}
\hline\hline
            &\multicolumn{1}{c}{Collaboration}&\multicolumn{1}{c}{Persisted Collaboration}\\
\hline
\textbf{Main effects}       &                     &                     \\
Cosine &       0.371\sym{***}&       0.409\sym{***}\\
            &     (0.027)         &     (0.042)         \\
[1em]
Proximity &       0.468\sym{***}&       0.847\sym{***}\\
           &     (0.019)         &     (0.037)         \\
[1em]
Jaccard &       0.867\sym{***}&       0.567\sym{***}\\
            &     (0.048)         &     (0.046)         \\
[1em]
Constant      &      -2.261\sym{***}&      -5.271\sym{***}\\
            &     (0.141)         &     (0.250)         \\
\hline
\textbf{Zero-inflation}     &                     &                     \\
Same country      &      -6.986\sym{***}&      -3.174\sym{***}\\
(dummy variable)            &     (0.260)         &     (0.151)         \\
[1em]
Log number of connections of region \textit{i} &      -11.96\sym{***}&      -0.522\sym{**} \\
            &     (1.180)         &     (0.204)         \\
[1em]
Log number of connections of region \textit{j} &      -13.89\sym{***}&      -0.801\sym{***}\\
            &     (0.750)         &     (0.133)         \\
[1em]
Log number of inventors in region \textit{i} &       3.877\sym{***}&      -3.371\sym{***}\\
            &     (0.683)         &     (0.381)         \\
[1em]
Log number of inventors in region \textit{j} &       11.12\sym{***}&       1.100\sym{***}\\
            &     (1.146)         &     (0.269)         \\
[1em]
Log population density in region \textit{i} &       0.370\sym{***}&      0.0454         \\
            &     (0.105)         &     (0.066)         \\
[1em]
Log population density in region \textit{j}&       0.710\sym{***}&       0.227\sym{***}\\
            &     (0.078)         &     (0.041)         \\
[1em]
Log gross value added in region \textit{i} &      -2.040\sym{***}&       0.345\sym{***}\\
            &     (0.119)         &     (0.131)         \\
[1em]
Log gross value added in region \textit{j}&      -5.137\sym{***}&      -0.657\sym{***}\\
            &     (0.340)         &     (0.095)         \\
[1em]
Constant      &       61.01         &       12.32\sym{***}\\
            &         (.)         &     (0.718)         \\
\hline
lnalpha     &                     &                     \\
Constant      &       2.220\sym{***}&       2.160\sym{***}\\
            &     (0.056)         &     (0.059)         \\
\hline
p           &           0         &           0         \\
Log lik.    &   -20,7048.8         &    -32,905.8         \\
\(N\)       &      872,235         &      872,235         \\
\hline\hline
\multicolumn{3}{l}{\footnotesize Standard errors in parentheses}\\
\multicolumn{3}{l}{\footnotesize \sym{*} \(p<0.10\), \sym{**} \(p<0.05\), \sym{***} \(p<0.01\)}\\
\end{tabular}
\end{table}

\begin{table}[ht]
\caption {Gravity models with interaction terms, zero-inflated negative binomial regression} \label{tab:SI4} 
\centering
\def\sym#1{\ifmmode^{#1}\else\(^{#1}\)\fi}
\begin{tabular}{l*{4}{c}}
\hline\hline
            &\multicolumn{1}{c}{Collaboration}&\multicolumn{1}{c}{Persisted Collaboration}&\multicolumn{1}{c}{Collaboration}&\multicolumn{1}{c}{Persisted Collaboration}\\
\hline
\textbf{Binary and interaction effects}       &                     &                     \\
Cosine &    1.420\sym{***}&   1.933\sym{***}&   0.094&       1.075\sym{**}\\
(reference category: low)   &    	 (0.049)         &     (0.070)         &     (0.098)         &     (0.424)         \\
[1em]
Proximity   &   1.574\sym{***}&   1.686\sym{***}&   0.394\sym{***}&       0.433\\
(reference category: low)      &	 (0.066) & 				(0.156)    &     (0.109)         &     	(0.434)         \\
[1em]
Jaccard  &    2.924\sym{***} &  2.931\sym{***}  &   1.340\sym{***}&       2.342\sym{***}\\
(reference category: low)      &  (0.114) &  (0.191)      &     (0.119)         &     (0.383)         \\
[1em]
Proxy $\times$ Jaccard    & &   &      1.067\sym{***}&      0.321\\
           & &  &     (0.141)         &     (0.476)         \\
[1em]
Proxy $\times$ Cosine    & &   &      0.212&      -0.444\\
           & &  &     (0.144)         &     (0.553)         \\
[1em]
Jaccard $\times$ Cosine    & &  &      1.023\sym{***}&      -0.631\\
         & &    &     (0.190)         &     (0.481)         \\
[1em]
Jaccard $\times$ Cosine $\times$ Proxy   & &   &      0.709\sym{***}&      2.034\sym{***}\\
          & &   &     (0.115)         &     (0.359)         \\
\hline
p      & 	0  & 0     &           0         &           0         \\
Log lik.  &	-225 777.2	 &  -37 693.3 &   -226 004.8          &    -37 645.09         \\
\(N\)    & 872 235 & 872 235   &      872 235         &      872 235         \\
\hline\hline
\multicolumn{3}{l}{\footnotesize Standard errors in parentheses}\\
\multicolumn{3}{l}{\footnotesize \sym{*} \(p<0.10\), \sym{**} \(p<0.05\), \sym{***} \(p<0.01\)}\\
\end{tabular}
\end{table}

\end{document}